\documentclass[9pt,twocolumn,twoside]{optica}

\setboolean{shortarticle}{false}
\setboolean{minireview}{false}

\usepackage[T1]{fontenc}       
\usepackage[utf8]{inputenc}
\usepackage[english]{babel}
\usepackage{amsmath,amsfonts,amsthm,amssymb}
\usepackage{mathtools}
\usepackage{graphicx}
\usepackage{float}
\usepackage[labelfont=default]{caption}
\usepackage{xcolor}
\usepackage{verbatim}
\usepackage{bm}
\usepackage{array}
\usepackage{gensymb}
\usepackage{empheq}

\newcommand{\ep}{\epsilon}

\title{Universal description of channel plasmons in two-dimensional materials}

\author[1,2,3,*]{P.~A.~D.~Gon\c{c}alves}
\author[3]{Sergey~I.~Bozhevolnyi}
\author[1,2,3]{N.~Asger~Mortensen}
\author[4]{N.~M.~R.~Peres}

\affil[1]{Department of Photonics Engineering, Technical University of Denmark, DK-2800 Kgs.~Lyngby, Denmark}
\affil[2]{Center for Nanostructured Graphene, Technical University of Denmark, DK-2800 Kgs.~Lyngby, Denmark}
\affil[3]{Centre for Nano Optics, University of Southern Denmark, Campusvej 55, DK-5230 Odense M, Denmark}
\affil[4]{Department and Center of Physics, and QuantaLab, University of Minho, PT-4710--057, Braga, Portugal}
\affil[*]{Corresponding author: padgo@fotonik.dtu.dk}

\ociscodes{(250.5403) Plasmonics; (240.6680) Surface plasmons; (230.7370) Waveguides; (160.4236) Nanomaterials.}

\begin{abstract}
Channeling surface plasmon-polaritons 
to control their propagation direction is 
of the utmost importance for future optoelectronic devices. 
Here, we develop an effective-index method to describe and characterize the properties of 2D material's channel plasmon-polaritons (CPPs) guided 
along a V-shaped channel. Focusing on the case of graphene, we derive a universal Schr{\"o}dinger-like equation from which one can 
determine the dispersion relation of graphene CPPs and corresponding field distributions at any given frequency, 
since they depend on the geometry of the structure alone. 
The results are then compared against more rigorous theories, having obtained a very good agreement. Our calculations show that CPPs in graphene and other 
2D materials are attractive candidates to achieve deep subwavelength waveguiding of light, holding potential 
as active components for the next generation of tunable photonic devices.
\end{abstract}

\begin{document}

\maketitle

%
\section{Introduction}

Surface plasmon-polaritons (SPPs)~\cite{MaradudinModern} have been considered a viable tool to tailor light-matter interactions at the 
nanoscale owing to their ability to squeeze electromagnetic fields below the diffraction limit~\cite{Gramotnev:2010,Barnes}. Such property 
makes SPP modes particularly attractive in the design of photonic components which rival their electronic counterparts in terms of miniaturization, 
while delivering larger bandwidths and operational speeds~\cite{Sci311}.
Among the vast number of applications in optoelectronics where SPPs may be put into use, one of the most prominent is subwavelength waveguiding 
of electromagnetic radiation. In this context, gap-SPPs (SPP modes sustained at dielectric gaps separating two metal surfaces) are appealing 
candidates due to their favorable balance between losses and field confinement~\cite{SergeyBook,Bozhevolnyi:2006,Moreno:06}. 
The width of the dielectric gap can either remain constant or vary in the direction perpendicular to it. An example of the latter configuration 
is a V-groove carved into a metallic substrate. Within this geometry, the corresponding propagating SPP modes are generally referred to as channel 
plasmon-polaritons (CPPs)~\cite{Ebbesen:2008,SmithRevCPP}. Contrarily to elementary (planar) dielectric-metal interfaces, where the field is confined only along 
one direction, CPP modes render a two-dimensional (2D) confinement of the electromagnetic field in the plane normal to its propagation (the latter 
being along the translational invariant direction).
As of today, a plethora of fundamental explorations~\cite{Bozhevolnyi:2005a,Sondergaard:2012,Raza:2014a} and
proof-of-concept experiments have been carried out demonstrating the usage of noble metal CPPs in plasmonic interferometers~\cite{Sergey_nat440,Haffner:2015}, waveguides~\cite{Nielsen:2008,PileAPL,telecom,Pile:04,Al_V}, 
ring-resonators~\cite{Sergey_nat440}, and for nanofocusing~\cite{nl9_nanoFocus}. In the context of quantum plasmonics~\cite{Bozhevolnyi:2017}, CPPs have also been explored for the control of quantum emitters~\cite{Bermudez-Urena:2015}. Indeed, such realizations unambiguously announce that the field is reaching maturity~\cite{SmithRevCPP}.

In parallel, the rise of graphene as a novel plasmonic medium has emerged only about half a decade ago~\cite{NatNano,GoncalvesPeres,Xiao2016,AbajoACSP}. 
In spite of the nanophotonics of 2D materials~\cite{2D_nphoton} being in its early stages, we have been witnessing an extraordinarily rapid progress in graphene plasmonics, fueled by the prospect of long-lived, gate-tunable graphene plasmons (GPs)
which yield large field confinements in the THz and mid-IR~\cite{GoncalvesPeres,Xiao2016,AbajoACSP}. At the time of writing, the majority of studies on graphene plasmons have dealt with nanostructured graphene (e.g., ribbons, disks, etc)~\cite{NatNano,Xiao2016,nphoton7,ACS7,NJP14,Zhu:2014}. Nonplanar graphene 
plasmons have recently gained interest~\cite{Kumar:2014,Christensen:2015,Riso:2015,Wang:2016}, 
while the investigation of graphene channel plasmons still remains largely unexplored. Only very recently, research on the plasmonic properties of 
graphene-covered triangular wedges and grooves has been conducted~\cite{Goncalves:16,Smirnova:2016,Liu:13}.

\begin{figure}[b!]
  \centering
    \includegraphics[width=0.475\textwidth]{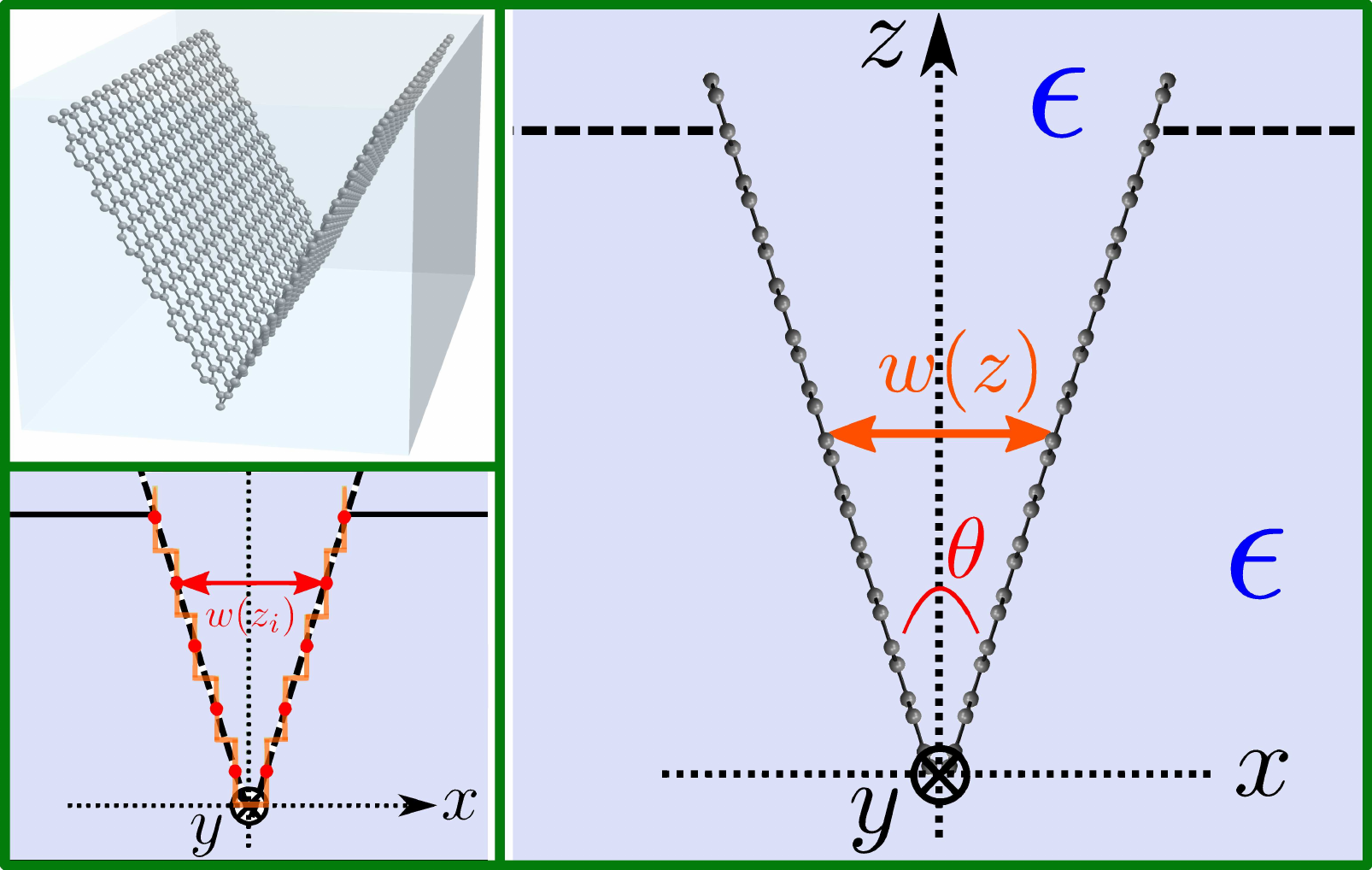}
      \caption{V-shaped groove carved in a dielectric substrate of relative permittivity $\ep$ onto which a 
      graphene sheet was deposited. After the deposition of graphene, the channel is then filled with the same dielectric material 
      as the underlying substrate. The channel width as a function of the $z$-coordinate follows $w(z)=2z\tan(\theta/2)$ for triangular 
      cross-sections.}\label{fig:system}
\end{figure}
In this paper, we derive a universal description to characterize channel plasmons guided along the apex of a triangularly shaped 
sheet of a conductive 2D material (e.g., graphene, 2D electron gas, doped 2D semiconductor transition metal dichalcogenides (TMDCs), etc); see Fig.~\ref{fig:system} with the 2D material exemplified by graphene. Our approach follows in part the effective-index framework previously applied to CPPs occurring in traditional 
metals~\cite{Bozhevolnyi:2006,Bozhevolnyi_oe17,Bozhevolnyi:09}. 
The concept behind the effective-index method (EIM) is that, at each height (with respect to the bottom of the channel), one can identify a one-dimensional (1D) 
dielectric/graphene/dielectric/graphene/dielectric (DGDGD) waveguide in which the core dielectric has a width given by $w(z)$. 
The combination of such 1D 
waveguide configurations---i.e., DGDGD slabs stacked together---can then be used to construct the whole two-dimensional channel 
waveguide akin to graphene CPPs; see Fig.~\ref{fig:system}. Here, we show that this conceptually intuitive picture can be used to determine the dispersion 
of CPPs and corresponding electric field distributions by solving a Schr{\"o}dinger-like equation whose eigenvalues depend only on the angle $\theta$ of the channel.
Finally, the results obtained using this unified description are then compared against the ones calculated using more rigorous theories~\cite{Goncalves:16}, having 
obtained a good agreement within the EIM regime.

%
%
%
%
\section{Methods}

We consider a V-shaped geometry making an angle $\theta$ as illustrated in Fig. \ref{fig:system}, where it is assumed henceforth that $\theta \ll 1$ 
in order for the EIM to be valid. Within this limit, one can make use of the transverse magnetic (TM) and electric (TE) representations. Therefore, we shall look for 
TM-like solutions in which the main component of the electric field lies along the $x$-axis. Recall that the underlying idea behind the 
EIM applied to the description of CPPs in V-grooves is to model the inner region as two coupled 1D waveguides. This foundational principle 
does not change when considering a dielectric V-channel covered with graphene (or any other 2D material). However, the corresponding equations will be quite different, asserting the natural differences between the two materials (i.e., 3D metal and 2D graphene). Therefore, we write the field as~\cite{Bozhevolnyi:2006,Bozhevolnyi_oe17,Bozhevolnyi:09} 
\begin{equation}
 E_x (\mathbf{r}) = X(x,z) Z(z) e^{i q y} \ , \label{def:E_x_spatial_distri_G}
\end{equation}
where $q$ denotes the propagation constant of the 2D material channel plasmon, and the coupled waveguide equation [stemming from the 
Helmholtz equation for the electric field (\ref{def:E_x_spatial_distri_G})] reads 
\begin{subequations}
\begin{align}
 \frac{\partial^2 X(x,z)}{\partial x^2} + \left[ \ep k_0^2 - \mathcal{Q}^2(z) \right] X(x,z) = 0 \ , \label{eq:subeq_Xxz_1_G} \\
 \frac{\partial^2 Z(z)}{\partial z^2} + \left[ \mathcal{Q}^2(z) - q^2 \right] Z(z) = 0 \ , \label{eq:subeq_Zz_2_G}
\end{align}
\label{subeqs:coupledEqs}%
\end{subequations}
with $k_0=\omega/c$, and where $\mathcal{Q}^2(z)$ acts as a separation constant; see Supplement 1. We note that in writing the above equations, 
we have neglected the $\partial^2 X/\partial z^2$ and $\partial X/\partial z$ derivatives, whose contribution is small in the regime 
where $\theta \ll 1$~\cite{Bozhevolnyi:09}. Notice that Eq.~(\ref{eq:subeq_Xxz_1_G}) is essentially an equation for a 1D waveguide, 
parameterized by $\mathcal{Q}(z)$. The specific form of $\mathcal{Q}(z)$ is, in general, non-trivial. Therefore, our strategy here 
consists in developing from the GP condition at a planar graphene double-interface with the same dielectric media---DGDGD---, which, 
for the mode with the sought-after symmetry, is given by~\cite{GoncalvesPeres}
\begin{align}
 1 + \coth\left[ \frac{w(z)}{2} \kappa_Q \right] 
  + \frac{i\sigma_{\mathrm{2D}}(\omega)}{\omega \ep \ep_0}\kappa_Q = 0\ , \label{eq:flat}
\end{align}
with $\kappa_Q = \sqrt{\mathcal{Q}^2(z) - \ep k_0^2}$, and where 
$w(z)= 2 z \tan(\theta/2) \approx \theta z$ is the gap-width as a function of the $z$-coordinate. Here, $\sigma_{\mathrm{2D}}(\omega)$ is the dynamical 
conductivity of graphene (or, more generally, any conducting 2D crystal). A closed-form expression for $\mathcal{Q}^2(z)$ may be obtained by 
expanding the previous equation for small widths \cite{note1}, 
which leads to \cite{note2}
\begin{equation}
 \mathcal{Q}^2(z) = \ep k_0^2 + \frac{\ep^ 2}{2 f_\sigma^ 2}
 \left[ 1 + \frac{4 f_\sigma}{\ep w(z)} + \sqrt{1 + \frac{8 f_\sigma}{\ep w(z)} } \right]\ , \label{eq:Q_z_Graphene}
\end{equation}
where $f_\sigma(\omega) = \mathrm{Im}\{\sigma_{\mathrm{2D}} \}/(\omega \ep_0)$. In the case of graphene, $f_\sigma(\omega) =4\alpha \hbar c \frac{E_F}{(\hbar \omega)^2}$ materializes the assumption that the conductivity is given by its Drude-like expression with negligible damping~\cite{GoncalvesPeres}, with $\alpha \simeq 1/137$ denoting the fine-structure constant. In possession of an explicit 
relation for $\mathcal{Q}^2(z)$, a solution for the coupled waveguide equations (\ref{subeqs:coupledEqs}) may be fetched. 
The solution to the first equation is trivial, simply being $X(x,z) = A \cosh( [\mathcal{Q}^2(z) - \ep k_0^2]^{1/2} x)$ in the inner region, 
where $A$ is a constant. The remaining differential equation (\ref{eq:subeq_Zz_2_G}) is somewhat less trivial. 
However, we can make such equation more affordable by defining the wavevector $q_0 \propto \omega^2$, denoting plasmons propagating in  
the planar 2D material~\cite{GoncalvesPeres}, and identifying $q_0=2\ep/f_\sigma$, while at the same time 
introducing the dimensionless variable $\zeta = \theta q_0 z$. Performing these transformations, 
we arrive at a dimensionless eigenvalue equation \cite{note3}
\begin{subequations}
\begin{equation}
- \theta^2\frac{\partial^2 Z(\zeta)}{\partial \zeta^2} + V(\zeta)  Z(\zeta) = \mathcal{E}_\theta Z(\zeta)\ ,
\end{equation}
resembling a Schr{\"o}dinger equation with an ``energy potential''
\begin{equation}
 V(\zeta) =- \frac{8 + \zeta +  \sqrt{\zeta^2+16\zeta}}{8\zeta}\ ,
\end{equation}
\label{eqs:dimensionless}%
\end{subequations}
where the 2D CPP dispersion relation $q^2=\ep k_0^2-\mathcal{E}_\theta  q_0^2$ is given in terms of the dimensionless eigenvalue $\mathcal{E}_\theta$. 
Interestingly, notice that $\mathcal{E}_\theta $ mimics 
an ``effective dielectric function'' that is entirely determined by the geometry---namely the angle $\theta$---of the channel. We stress that the solution of Eq. (\ref{eqs:dimensionless}) 
yields the eigenfunctions and propagation constants of the CPP modes, while holding the advantage of portraying a clearer picture of the physics of the problem which is epitomized by the universal scaling behavior for the CPP wavevector depending entirely on the geometry of the V-channel. 

We bring to the reader's attention that the above differential equation for $Z(z)$ is substantially more intricate in our present case with a 2D material than 
it is for conventional CPPs in metallic V-grooves, where a simple differential equation and corresponding \emph{analytic} solution can be 
straightforwardly derived~\cite{Bozhevolnyi:09}. In what follows, we solve the differential equation~(\ref{eqs:dimensionless}) 
numerically using the shooting method~\cite{nrec}.

\section{Results and discussion}

\begin{figure}[b!]
  \centering
    \includegraphics[width=0.35\textwidth]{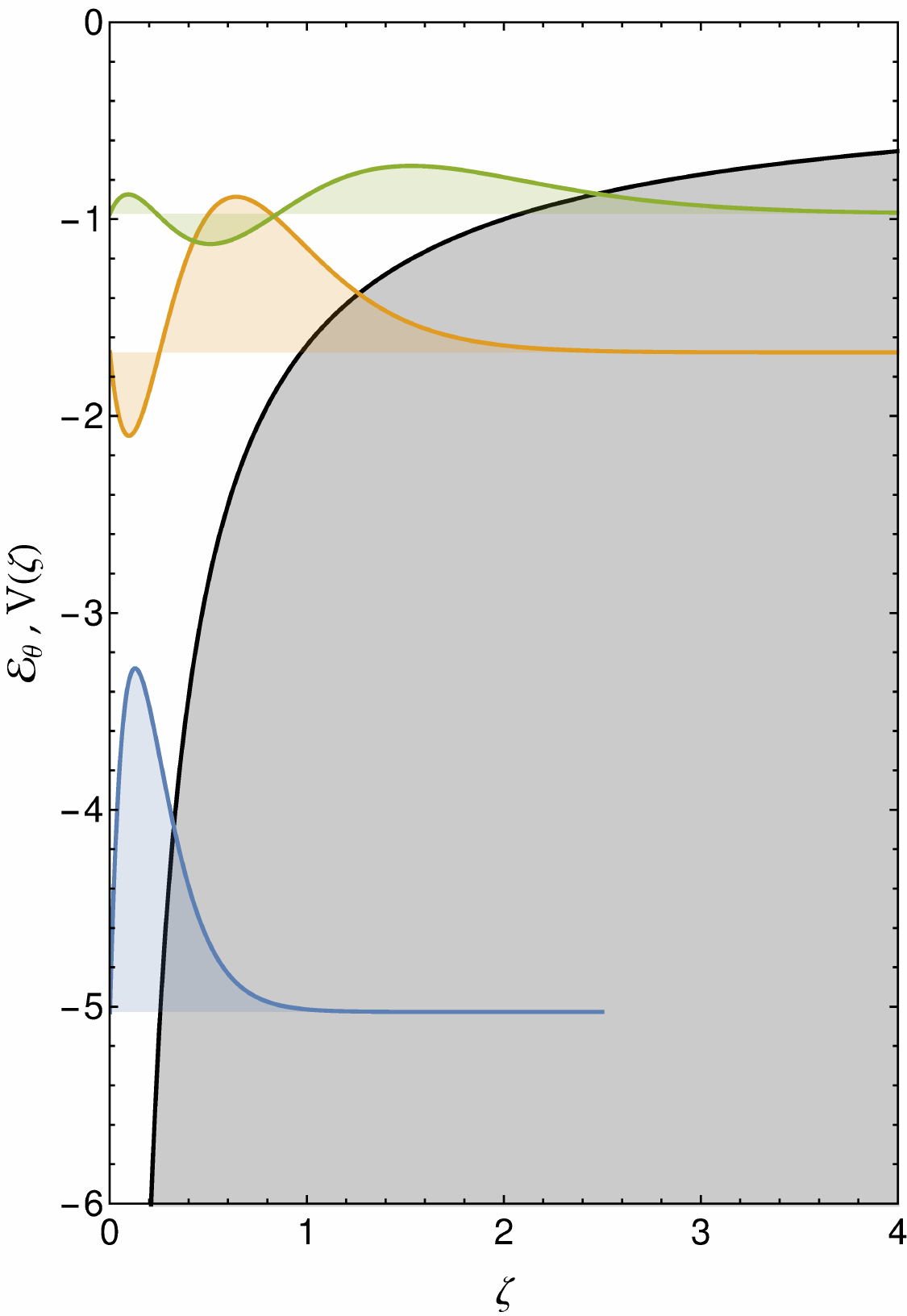}
      \caption{Illustration of the field distributions along the height of the channel, $Z_n(\zeta)$, akin to the first three 2D CPPs modes in a triangular 
      structure with $\theta=15\degree$. We depict the eigenfunctions in an energy diagram along with the potential $V(\zeta)$ (black line), as in typical
      quantum mechanical problems. The vertical axis for each $Z_n$ (in a.u.) starts at the position of the corresponding eigenvalue, $\mathcal{E}_\theta^{(n)}$. 
      }\label{fig:Zn_z_G}
\end{figure}
The eigenfunctions $Z(z)$ resulting from the numerical solution of Eq.~(\ref{eqs:dimensionless}) are depicted in Fig.~\ref{fig:Zn_z_G}, 
for 2D CPP modes in a V-shaped channel with $\theta=15\degree$. As portrayed in the figure, these behave like bound modes trapped in a potential well, 
even displaying features familiar to the ones encountered in typical quantum mechanics problems, 
such as the decrease in energy difference as the mode order increases and the number of nodes given by $n-1$ (where $n=1,2,...$ labels the mode order). 
Moreover, Fig. \ref{fig:Zn_z_G} clearly shows that the fundamental CPP mode is highly confined within the bottom of the groove (i.e., near the apex of the channel), 
while the successively higher order modes tend to be concomitantly more spread along the vertical direction of the channel 
[i.e., the ($\zeta$-) $z$-direction].

\begin{figure}[t]
  \centering
    \includegraphics[width=0.45\textwidth]{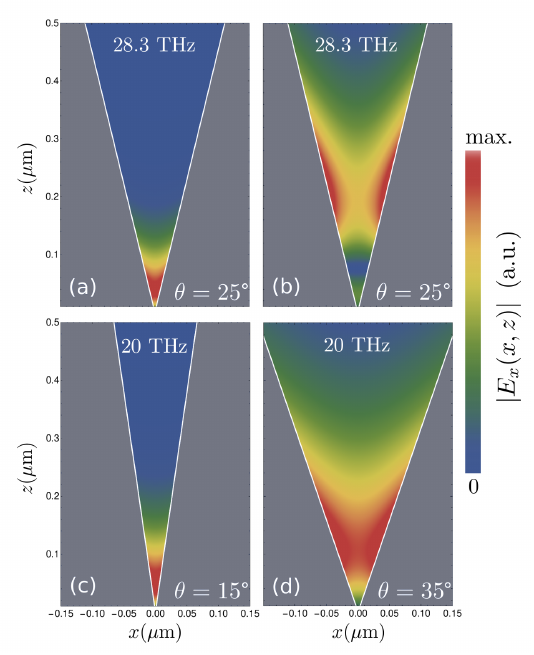}
      \caption{Two-dimensional distributions of the electric field magnitude, $| E_x^{(n)}(x,z) |$, for CPPs in several triangular geometries: fundamental (a), (c)--(d), and 
      second-order (b) guided plasmonic modes. The resonant frequencies and opening angles are indicated in each panel. The field is depicted only in the inner region. 
      Parameters: $E_F=0.5$\,eV and $\ep=2.1$.}\label{fig:2D}
\end{figure}
Having computed the eigenvalues and field profiles along $z$-coordinate, one may readily construct the 2D electric field distributions, $E_x^{(n)}(x,z)$, corresponding to the aforementioned CPP modes using Eq.~(\ref{def:E_x_spatial_distri_G}). The outcome of such procedure is shown in Fig.~ \ref{fig:2D}, for V-shaped 
graphene with different opening angles, and excited at different resonant frequencies. The top panels illustrate the modal distributions for the 
fundamental [Fig.~\ref{fig:2D}(a)] and second-order [Fig.~\ref{fig:2D}(b)] CPP modes, whose main features echo our above description of Fig.~\ref{fig:Zn_z_G}, namely the higher level of localization achieved by the fundamental mode when compared with its leading order ones. In addition, 
the influence of the groove angle is reflected in the two bottom panels 
of the figure, with smaller opening angles yielding correspondingly higher field confinements near the apex of the channel. This behavior is consequence 
of the stronger ``confining potential'' for CPPs in sharper V-structures. Furthermore, let us stress that we expect that the propagation constants and corresponding electric 
\begin{figure*}[h]
  \centering
    \includegraphics[width=0.95\textwidth]{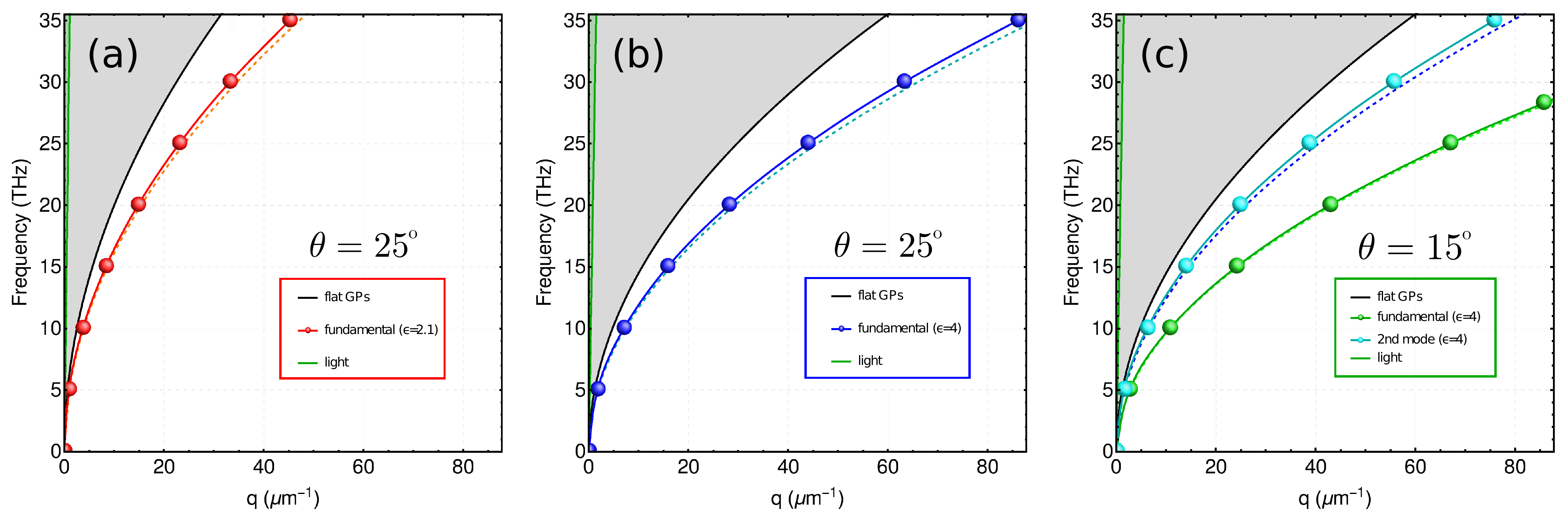}
      \caption{Dispersion relation of guided graphene CPPs sustained in a V-shaped graphene configuration embedded in 
      homogeneous dielectric environments with different dielectric constants $\ep$, for two different opening angles; see figure's labels. 
      As an eye-guide, the solid black lines indicate the dispersion of GPs in a corresponding flat interface. We assume a Fermi energy 
      of $E_F=0.5$ eV in the calculations. For comparison, the dashed lines show the dispersion relation of CPPs obtained using a more 
      rigorous theory \cite{Goncalves:16}.}\label{fig:dispersion}
\end{figure*}
field profiles to remain accurate even for finite-sized channels, as long as the truncation of the V-shaped structure occurs at sufficiently large distances away from the apex 
(i.e., at $z$-coordinates where the field is essentially zero). In this regard, the computed field patterns may be used to estimate the required channel depth/height a device based on 2D CPPs should have.

\begin{figure}[h]
  \centering
    \includegraphics[width=0.425\textwidth]{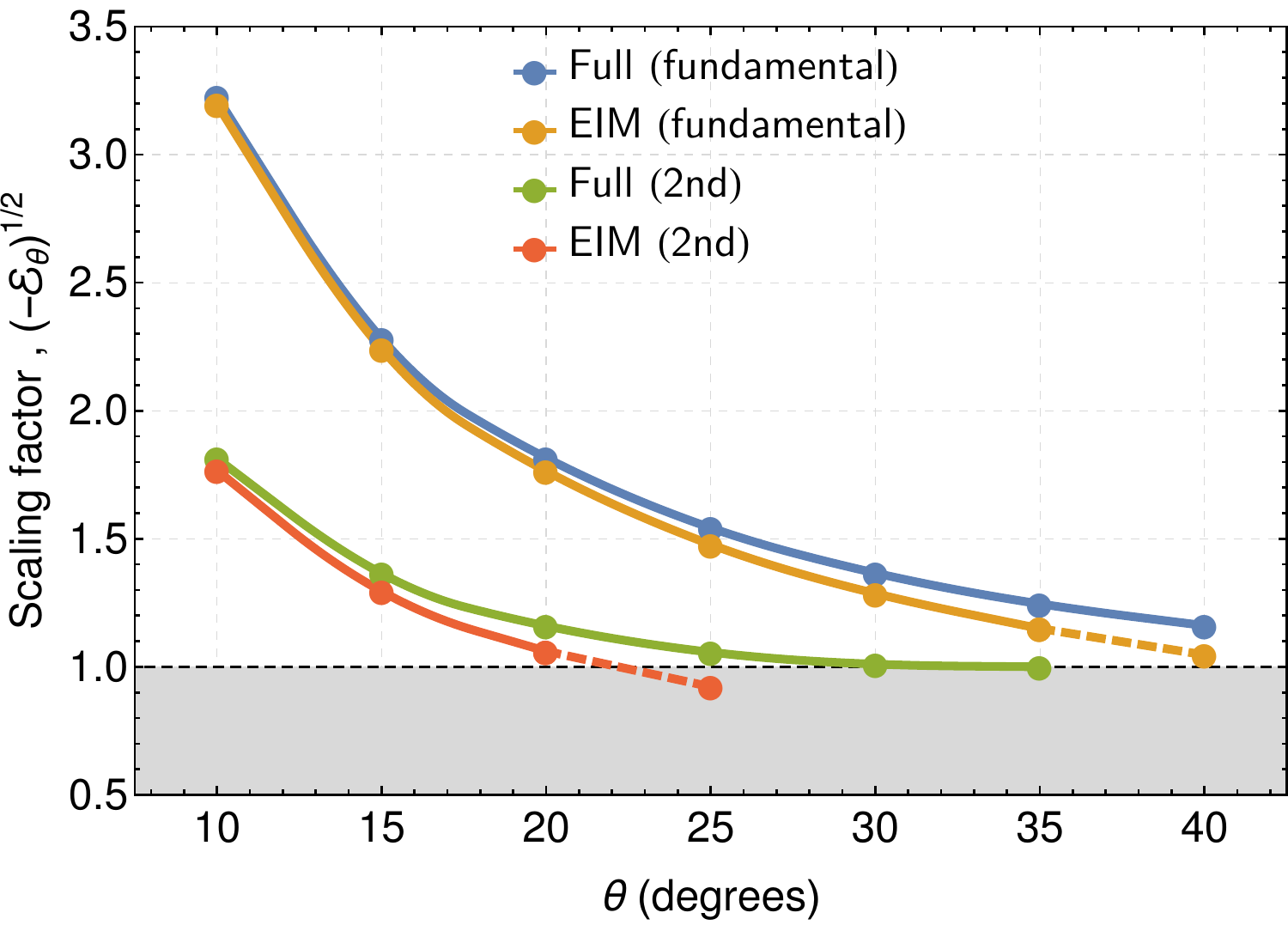}
      \caption{Universal description of 2D CPPs, namely $\sqrt{-\mathcal{E}_\theta} \simeq q/q_0$ as a function of the angle $\theta$ (EIM). 
      The results obtained using a rigorous theory~\cite{Goncalves:16} (Full) are also plotted for comparison. The dashed lines indicate the point 
      where the EIM model seems to surpass its regime of validity. The EIM curves follow a simple analytical expression of the 
      form $a_n+b_n \theta^{-1}$ with $a_n=\{0.33,0.36\}$ and $b_n=\{28.7,14.1\}$ (with $n$ denoting the mode order); see Supplement 1.}\label{fig:E_vs_theta}
\end{figure}

A characteristic of particular physical importance, both from a fundamental perspective and also from a device-engineering viewpoint, is the dispersion relation exhibited by the graphene CPPs. The 
spectrum of CPPs guided along triangular channels stems from the solution of Eq.~(\ref{eqs:dimensionless}), 
which allows the retrieval of the propagation constant in terms of its planar graphene plasmon counterpart. Thus, with a single calculation one may 
construct the entire plasmon dispersion in the system of interest, for any frequency or any other parameter, since the solution depends uniquely on the angle $\theta$ of the structure.
In Fig.~\ref{fig:dispersion} we show the obtained dispersion relations for CPP in representative triangular channels. Figures~\ref{fig:dispersion}(a)--(b) exhibit the 
spectrum for the fundamental CPP in two dielectric environments with different 
relative permittivities, keeping the angle $\theta = 25\degree$ constant. As an eye-guide, we have also plotted the dispersion of graphene plasmons in flat graphene, along 
with the dispersion of light propagating freely inside the dielectric. It is apparent from the figure that the dispersion curve akin to graphene CPPs lies 
to the right of their corresponding flat GPs. This is indicative of the higher amount of field 
localization attained by the former when compared with the latter. 
The degree of subwavelength localization of the electromagnetic field is 
even more dramatic for smaller angles, as a comparison between Fig. \ref{fig:dispersion}(b) and Fig. \ref{fig:dispersion}(c)  plainly shows.
Additionally, in order to gauge the level of fidelity of our model we have also superimposed the dispersion of the CPP modes 
obtained using a more rigorous technique \cite{Goncalves:16} (dashed lines in Fig.~ \ref{fig:dispersion}). 
Performing a comparison of results, a very good agreement can be observed regarding the dispersion of the fundamental CPP mode in Fig.~\ref{fig:dispersion}(a)--(b), 
and both the fundamental and second-order mode in Fig. \ref{fig:dispersion}(c). This 
clearly illustrates that, despite its inherent simplicity, the EIM framework can still provide consistent results, thus being 
a very valuable tool in judging the main characteristics of 2D material's channel plasmons. 
Unfortunately, the EIM does not capture the higher-order modes in the case of $\theta=25\degree$ (not shown); the justification for this 
lies in the breakage of the small-argument expansion performed in the cotangent figuring in Eq.~(\ref{eq:flat}), as the 
field distributions arising from these (higher-order) modes extend to regions relatively far away from the apex [cf. Fig.~\ref{fig:2D}(b)]. 
Nevertheless, note that for smaller angles (e.g., $\theta=15\degree$) the second-order mode is already well described by the EIM owing to 
the smaller gap. A more all-encompassing perspective can be laid out by taking advantage of the universal scaling of the graphene (or any 2D material) CPPs spectrum 
with the angle $\theta$, given by the eigenvalues $\mathcal{E}_\theta$. Figure \ref{fig:E_vs_theta} shows the scaling factor, 
$q/q_0 \simeq \sqrt{-\mathcal{E}_\theta}$ (since $q,q_0 \gg k_0$) as a function of the opening angle of the system.
The plot demonstrates quite remarkably the ability of our model to correctly determine the behavior of 2D channel plasmons. In 
the case of the fundamental mode, the agreement is maintained even beyond the small angle ($\theta \ll 1$) regime. Naturally, 
and as already anticipated above, the description of the higher-order modes is limited to smaller angles. Figure \ref{fig:E_vs_theta} 
plays the role of an epilogue: it clearly displays the universal scaling law for the 2D CPP's propagation constant as a function 
of the angle---valid independently of the frequency, 2D conductivity, or dielectric constant---, while at the same time outlining 
the regime of validity of the model. Moreover, we have found that the curves plotted in the figure follow a simple analytical 
expression, in the form of $a+b \theta^{-1}$, to which we have fitted the constants $a$ and $b$; cf. Supplement 1. Such feature may 
be of extreme relevance in the experimental design of graphene-based CPPs, functioning as a ``ruler'' or ``recipe'' 
to fabricate graphene CPPs with tailored plasmonic properties. In fact, this relation is entirely general irrespective of the 2D 
material which is employed, since the specific properties of each material are all encapsulated within the flat $q_0$, 
which is known; check Supplement 1. Therefore, although we have mainly focused on graphene throughout this manuscript, our results can be made
completely general, as in Fig. \ref{fig:E_vs_theta}, for any plasmonic 2D medium.

%
%
%
\section{Conclusions}

In conclusion, we have developed a universal framework to characterize plasmonic excitations propagating along the apex 
of a V-shaped graphene sheet forming a 1D channel. Using the effective-index method applied to the case of graphene, 
we derived the dispersion relation for graphene CPPs and their corresponding electric field distributions. As our 
calculations demonstrate, the EIM not only endows us with an intuitive picture of the physics governing these guided modes, 
but it can also provide results consistent with the ones obtained using more elaborated theories, in particular for the fundamental 
channel plasmon mode (which in any case should be the most relevant from an applications perspective owing to its ability to reach deeper subwavelength regimes). 
This is courtesy of a simple yet powerful universal equation and corresponding scaling law that depends only on the system's geometry (namely the angle $\theta$). 
Therefore, we believe our work may be applied as a first step---since the solution is essentially obtained \emph{instantaneously}---in the design of novel 
devices for extreme subwavelength waveguiding of electromagnetic radiation~\cite{Bozhevolnyi:2017} based either on graphene channel plasmon-polaritons 
or any other plasmon-supporting 2D material belonging to the now vast ``2D library'' of atomically thin materials~\cite{vdW_Nature}.

Furthermore, in principle, one could envision the inclusion of a gain medium for loss-compensation in order to obtain longer propagation lengths. The effect of gain media could be incorporated in our model by replacing $\epsilon$ by an adequate dielectric function $\epsilon(\omega)$, so that it balances the Ohmic losses of the 2D material \cite{xiao2010loss}. Note, however, that this is possible only provided 
that we are below the lasing threshold and for $\mathcal{Q}^2(z) \gg \epsilon k^2_0$; cf. Eq. \ref{eq:Q_z_Graphene}. Naturally, a rigorous and complete description of gain can only be achieved using a 
quantum optical framework, see for instance Ref.~\cite{PRL_Wubs}.

Finally, as a last remark, we note that our notion of a 2D material may in principle also include ultra-thin metallic films. For instance, 
a planar Drude-metal film of finite, but small thickness $d$, has a symmetric mode with a low-frequency dependence of the form ~\cite{Raza:2013} $q_0\propto\omega^2/d$, 
thus being analogous to our generic 2D materials. In this way, CPPs in ultra-thin wedge/groove geometries~\cite{Lee:11,Chen:2016} can be also covered by the universality outlined above.

%

\section*{Funding Information}
The Center for Nanostructured Graphene is supported by the Danish National Research Foundation (DNRF103). N.M.R.P. acknowledges 
funding from the European Commission within the project “Graphene-Driven Revolutions 
in ICT and Beyond” (ref. no. 696656) and the Portuguese Foundation for Science and Technology (FCT) in the 
framework of the Strategic Financing UID/FIS/04650/2013. S.I.B. acknowledges financial support by the 
European Research Council, Grant 341054 (PLAQNAP). N.A.M. is a Villum Investigator supported by Villum Fonden.


\bigskip \noindent See \href{link}{Supplement 1} for supporting content.

\bibliography{refs_EIM_Graphene}

\begin{thebibliography}{10}
\bibitem{note2}
 We remark that if the dielectric material within the V-shaped region differs
  from the cladding dielectric media, it is not possible to derive a simple,
  closed-form expression for $\mathcal{Q}^2(z)$. For this reason, here we
  consider only the case of a homogeneous dielectric environment.
\bibitem{note3}
 In writing this equation we have explicitly approximated the width (for small
  angles) as $w(z) \approx \theta z$, for the sake of clarity. The expression
  using the \emph{exact} form is recoverable upon the replacement $\theta
  \rightarrow 2 \tan(\theta/2)$ (we have used the \emph{exact} expression in
  all our results).
\end{thebibliography}


\end{document}